\documentclass[twocolumn,showpacs,superscriptaddress,prb,aps,floatfix]{revtex4-2}
\usepackage{graphicx}
\usepackage{rotating}
\usepackage{amsmath}
\usepackage[usenames,dvipsnames]{color}
\usepackage{float}
\usepackage{rotating}
\usepackage{booktabs}
\usepackage{multirow}
\usepackage{longtable}
\usepackage[colorlinks,linkcolor=blue,citecolor=magenta]{hyperref}
\usepackage[T1]{fontenc}
\usepackage{bm}
\usepackage{upgreek}
\usepackage{comment}
\usepackage{amssymb}

\makeatletter

\makeatletter

\hfuzz=\maxdimen
\begin{document}

\title{Microscopic Origin of the Ultralow Lattice Thermal Conductivity in Vacancy-Ordered Halide Double Perovskites Cs$_2BX_6$ ($B$ = Zr, Pd, Sn, Te, Hf, and Pt; $X$= Cl, Br, and I)}

\author{Lingzhi Cao}
\affiliation{State Key Laboratory for Advanced Metals and Materials, Beijing Key Laboratory for Magneto-Photoelectrical Composite and Interface Science, School of Mathematics and Physics, University of Science and Technology Beijing, Beijing 100083, China}

\author{Yateng Wang}
\affiliation{State Key Laboratory for Advanced Metals and Materials, Beijing Key Laboratory for Magneto-Photoelectrical Composite and Interface Science, School of Mathematics and Physics, University of Science and Technology Beijing, Beijing 100083, China}

\author{Zhonghao Xia}
\affiliation{State Key Laboratory for Advanced Metals and Materials, Beijing Key Laboratory for Magneto-Photoelectrical Composite and Interface Science, School of Mathematics and Physics, University of Science and Technology Beijing, Beijing 100083, China}

\author{Jiangang He}
\email{jghe2021@ustb.edu.cn}
\affiliation{State Key Laboratory for Advanced Metals and Materials, Beijing Key Laboratory for Magneto-Photoelectrical Composite and Interface Science, School of Mathematics and Physics, University of Science and Technology Beijing, Beijing 100083, China}

\date{\today}


\begin{abstract}
Lead-free halide vacancy-ordered double perovskites Cs$_2BX_6$ have recently attracted significant attention due to their intrinsically ultralow lattice thermal conductivity ($\kappa_{\mathrm{L}}$), which is highly desirable for thermal insulation and thermoelectric applications. In this work, we systematically investigate the anharmonic lattice dynamics and thermal transport properties of Cs$_2BX_6$ ($B$ = Zr, Pd, Sn, Te, Hf, and Pt; $X$ = Cl, Br, and I) using state-of-the-art first-principles calculations, based on a unified theory of thermal transport for crystals and glasses. All studied compounds are found to exhibit ultralow $\kappa_{\mathrm{L}}$ below 1.0~W\,m$^{-1}$\,K$^{-1}$ at room temperature and large derivation from the conventional $T^{-1}$ temperature dependence. Our analysis combining with machine-learning approach show that low sound velocities (1100 -- 1600~m\,s$^{-1}$), which originates from the intrinsically weak chemical bonding, play a crucial role in suppressing heat transport of the most compounds, instead of the strong scattering of rattling phonon modes expected from the large void in the structure. Furthermore, the influence of $B$ and $X$-site elements on phonon dispersion, anharmonicity, and scattering phase space is clarified. Our results provide microscopic insights into the origin of ultralow $\kappa_{\mathrm{L}}$ in Cs$_2BX_6$ and offer guiding principles for the rational design of halide-based materials with tailored thermal transport properties.
\end{abstract}


\maketitle

\section{Introduction}
The $\kappa_{\mathrm{L}}$ is a fundamental property of solids and plays a critical role in determining the performance of optoelectronic and thermoelectric devices. Materials with extremely low or high $\kappa_{\mathrm{L}}$ are also essential for applications such as thermoelectric energy conversion, thermal management, and thermal barrier coatings~\cite{padture2002thermal,science5895,qian2021phonon}. Within the framework of kinetic theory~\cite{tritt2005thermal}, the $\kappa_{\mathrm{L}}$ of crystalline solids can be expressed as $\kappa_{\mathrm{L}} = \frac{1}{3} C_{\mathrm{V}} \nu_{\mathrm{g}}^{2} \tau$, where $C_{\mathrm{V}}$ is the heat capacity, $\nu_{\mathrm{g}}$ is the phonon group velocity, and $\tau$ is the phonon relaxation time. Accordingly, ultralow $\kappa_{\mathrm{L}}$ typically arises from either a small $\nu_{\mathrm{g}}$ or a short $\tau$. Strong lattice anharmonicity generally leads to enhanced phonon--phonon scattering (large $1/\tau$), which can originate from rattling phonon modes~\cite{annurev:/content/journals/10.1146/annurev.matsci.29.1.89,PhysRevB.91.144304,C1JM11754H,christensen2008avoided}, cations with stereochemically active lone-pair electrons~\cite{C2EE23391F,PhysRevLett.107.235901}, or resonant bonding~\cite{lee2014resonant}. In contrast, a low phonon group velocity is commonly realized in compounds with weak chemical bonding (small bond stiffness $K$) and large atomic masses ($M$), following the approximate relation $\nu_{\mathrm{g}} \propto \sqrt{K/M}$~\cite{https://doi.org/10.1002/adfm.202108532,https://doi.org/10.1002/advs.202417292,WANG2020100180}.

The vacancy-ordered double perovskite $A_{2}BX_{6}$ is a structural derivative of the conventional $B$-site rock-salt–ordered double perovskite ($A_{2}BB^{\prime}X_{6}$), in which $B^{\prime}$ sublattice is replaced by ordered vacancies~\cite{doi:10.1021/acs.chemmater.8b05036}. Within this materials family, Cs$_{2}BX_{6}$ compounds have attracted considerable interest because of their unique chemical composition, characterized by the occupation of half of the $B$-site positions by tetravalent cations. Recent studies have highlighted the strong potential of Cs$_{2}BX_{6}$ materials for optoelectronic and thermoelectric applications, owing to their favorable optical and electrical properties, low thermal conductivity, and excellent stability under ambient conditions~\cite{Zhou2021,HUSSAIN202021378,Faizan2021,Sreedevi2021, HU2022236, Sajjad2020, Zeng2022, Bhui2022, Cai2017, https://doi.org/10.1002/anie.202016185, https://doi.org/10.1002/anie.202000175}. For instance, Cs$_2$SnI$_6$ is a native $n$-type semiconductor with a high electron mobility of 310~cm$^{2}$V$^{-1}$s$^{-1}$~\cite{doi:10.1021/acs.chemmater.8b05036}, and it also exhibits an exceptionally low $\kappa_{\mathrm{L}}$~\cite{Bhui2022}. In addition, Cs$_2$ZrCl$_6$ and Cs$_2$HfCl$_6$ have been identified as promising scintillator materials due to their excellent energy resolution; their room-temperature scintillation light yields under $^{137}$Cs $\gamma$-ray excitation reach 33900 and 24800 photons per MeV, respectively~\cite{D2DT00223J}.

In recent years, significant progress has been made in understanding the lattice thermal transport of vacancy-ordered double perovskites $A_{2}BX_{6}$. Zeng \textit{et al.}~\cite{doi:10.1021/acs.jpclett.2c02350} reported that Cs$_2$SnBr$_6$ exhibits an extraordinarily low $\kappa_{\mathrm{L}}$ and glass-like heat transport behavior, originating from the interplay between particle-like and wave-like phonon transport via interbranch tunneling, as revealed by first-principles calculations. Cheng \textit{et al.}~\cite{PhysRevB.109.054305} investigated the influence of mechanical strain on the anharmonic lattice dynamics of Rb$_2$SnBr$_6$ and Cs$_2$SnBr$_6$ by combining first-principles calculations with the unified theory of thermal transport in crystals and glasses~\cite{Simoncelli2019}. They showed that low-frequency optical phonons undergo pronounced hardening with increasing temperature, particularly in Rb$_2$SnBr$_6$, which hosts extremely soft low-lying optical modes that strongly suppress $\kappa_{\mathrm{L}}$. Bhui \textit{et al.}~\cite{bhui2022intrinsically} examined the thermoelectric properties of Cs$_2$SnI$_6$ using a combination of experiments and first-principles calculations, and demonstrated an ultralow $\kappa_{\mathrm{L}}$ of 0.29~W\,m$^{-1}$K$^{-1}$ at room temperature, attributed to its low $\nu_{\mathrm{g}}$ and short $\tau$. Furthermore, Li \textit{et al.}~\cite{doi:10.1021/jacs.1c11887} performed a high-throughput screening of materials with rattling phonon modes based on the distance to nearest neighbors, followed by empirical evaluations of $\kappa_{\mathrm{L}}$. They identified 1171 semiconductors with ultralow $\kappa_{\mathrm{L}}$, including 23 vacancy-ordered double perovskites $A_{2}BX_{6}$. Despite these advances, a systematic understanding of how $\kappa_{\mathrm{L}}$ evolves with $B$- and $X$-site chemistry across the Cs$_2BX_6$ family, as well as the microscopic mechanisms governing their ultralow $\kappa_{\mathrm{L}}$ at the atomic scale, remains lacking.

In this work, we present a comprehensive study of lattice thermal transport in vacancy-ordered double perovskites Cs$_2BX_6$ ($B$ = Zr, Pd, Sn, Te, and Hf; $X$ = Cl, Br, and I) using first-principles calculations within the unified theory of thermal transport for crystals and glasses~\cite{simoncelli2019unified}. Our results reveal that all investigated compounds exhibit ultralow $\kappa_{\mathrm{L}}$ and pronounced deviations from the conventional $T^{-1}$ temperature dependence expected for crystalline solids. Assisted by machine-learning analysis, we demonstrate that a low phonon group velocity, arising from intrinsically weak chemical bonding in these halide perovskites, is the primary factor suppressing lattice thermal transport across the entire series. In contrast, strong three-phonon scattering plays a dominant role only in Cs$_2$SnI$_6$, where flat phonon branches associated with iodine atoms significantly enhance anharmonic interactions. These findings provide fundamental insights into the intrinsic thermal transport mechanisms of vacancy-ordered double perovskites and offer guidance for their rational optimization in thermoelectric and other functional applications.

\section{COMPUTATION  DETAILS}

All density functional theory (DFT) calculations were carried out using the Vienna \textit{Ab initio} Simulation Package (\textsc{VASP})~\cite{PhysRevB.54.11169,vasp2}. The interactions between valence and core electrons were described using the projector augmented-wave (PAW) method~\cite{PhysRevB.50.17953,PAW2}. We employed the PBEsol~\cite{PhysRevLett.100.136406} exchange--correlation functional, which has been shown to perform well in $\kappa_{\mathrm{L}}$ calculations~\cite{PhysRevB.110.035205}. A plane-wave basis set with a kinetic-energy cutoff of 520~eV and a Monkhorst--Pack $k$-point mesh of $12 \times 12 \times 12$ were used. All crystal structures were fully relaxed until the residual force on each atom was less than 0.01~eV/\AA, and the total-energy convergence criterion was set to $10^{-5}$~eV. Crystal orbital Hamilton population (COHP) analyses were performed using the \textsc{LOBSTER} code~\cite{nelson2020lobster}. The bulk modulus ($B_0$), shear modulus ($G$), and average sound velocity ($\nu_{\mathrm{g}}$) were derived from elastic constants calculated using the finite-difference method.

The harmonic force constants (FCs) were computed using the finite-displacement method as implemented in the \textsc{Phonopy} package~\cite{PhysRevB.77.144112,TOGO20151}, employing a $4 \times 4 \times 4$ supercell containing 576 atoms. To improve the accuracy of the force calculations, the reciprocal-space projection scheme (LREAL = FALSE) and a stringent electronic convergence threshold of $10^{-8}$~eV were adopted. Higher-order force constants were extracted using the compressive sensing lattice dynamics (CSLD) method~\cite{PhysRevLett.113.185501}. Renormalization of the second-order force constants was performed using the self-consistent phonon (SCPH) method~\cite{PhysRevLett.106.165501,PhysRevLett.120.105901}, as implemented in the \textsc{ALAMODE} package~\cite{Tadano_2014}. Lattice thermal transport properties were obtained by iteratively solving the Peierls--Boltzmann transport equation using the \textsc{FourPhonon} package~\cite{HAN2022108179}. Uniform $16 \times 16 \times 16$ $q$-point meshes were used for $\kappa_{\mathrm{L}}$ calculations within the SCPH plus three-phonon (SCPH+3ph, $\kappa_{\mathrm{3ph}}^{\mathrm{SCPH}}$) framework, while the same $16 \times 16 \times 16$ $q$-point meshes were employed for calculations including both three- and four-phonon scattering processes (SCPH+3,4ph, $\kappa_{\mathrm{3,4ph}}^{\mathrm{SCPH}}$). The four-phonon scattering processes were accelerated using a sampling-based approach~\cite{Guo2024SamplingacceleratedPO}. The contribution of the coherent term to the lattice thermal conductivity ($\kappa_{\mathrm{L}}^{\mathrm{C}}$), within the unified thermal transport theory framework~\cite{simoncelli2019unified}, was calculated using our in-house code.

\section{RESULTS AND DISCUSSION}

\begin{figure}[tph!]
	\includegraphics[clip,width=1.0\linewidth]{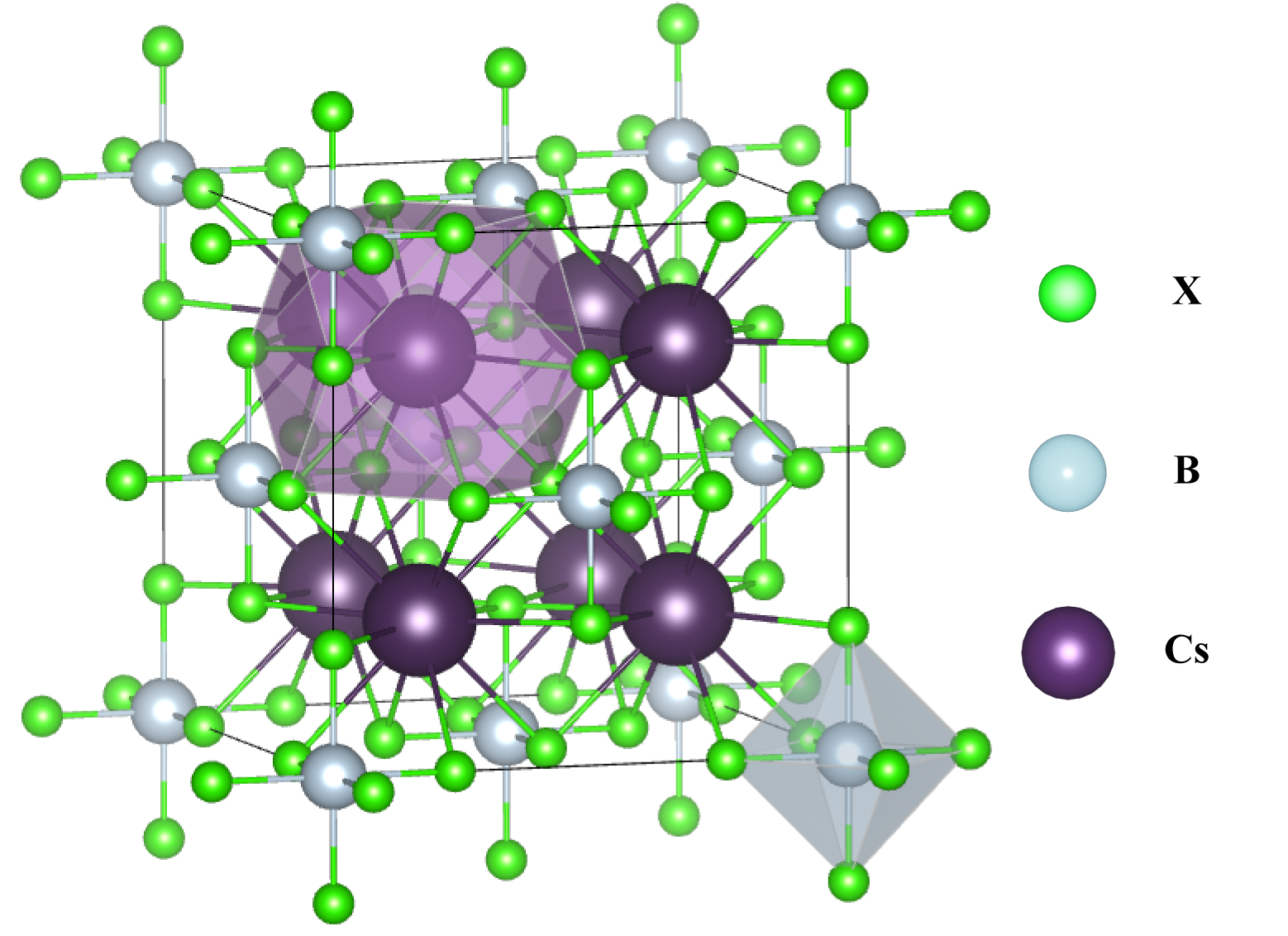}
	\caption{Crystal structure of the vacancy-ordered perovskite Cs$_2BX_6$. The 12-fold coordinated Cs atoms and the octahedrally coordinated $B$ atoms are highlighted using polyhedra.}
	\label{fig:st}
\end{figure}

\subsection{Structure of Cs$_2BX_6$ compounds}
Similar to perovskites $ABX_3$, vacancy-ordered double perovskites $A_2BX_6$ can undergo various octahedral distortions depending on the radius ratio of the $A$- and $B$-site cations. This ratio is commonly characterized by the Goldschmidt tolerance factor $\tau$~\cite{goldschmidt1926gesetze,doi:10.1021/acs.chemmater.8b01549}. Compounds with $\tau \approx 1.0$ typically adopt the cubic structure (space group $Fm\bar{3}m$, No.~225). As in conventional perovskites, the electronic structure and physical properties of $A_2BX_6$ are primarily determined by the $B$-site cation~\cite{doi:10.1021/acs.chemmater.8b01549}. However, because the physical properties of perovskites are strongly correlated with the $B$--$X$--$B$ bond angle, which depends on the magnitude of the octahedral distortion~\cite{doi:10.1021/acs.chemmater.8b01549}, the radii of the $A$- and $B$-site cations control the amplitude of octahedral distortions and can therefore indirectly influence the physical properties of $A_2BX_6$ compounds.

Experimentally, the $A$, $B$, and $X$ sites are typically occupied by alkali metals, tetravalent transition-metal or main-group elements, and halide elements, respectively~\cite{doi:10.1021/acs.chemmater.8b05036}. In this work, we focus on Cs-based compounds Cs$_2BX_6$ ($B$ = Zr, Pd, Sn, Te, Hf, and Pt; $X$ = Cl, Br, and I), as they generally exhibit large $\tau$ values and are therefore more likely to crystallize in the cubic structure at room temperature. Among the $B$-site elements, Zr and Hf are early transition metals (Zr$^{4+}$ and Hf$^{4+}$ have a $d^{0}$ configuration), Pd and Pt are late transition metals (Pd$^{4+}$ and Pt$^{4+}$ have a $d^{6}$ configuration, corresponding to $t_{2g}^{6}e_g^{0}$ under an octahedral crystal field), Sn is a post-transition metal (Sn$^{4+}$ has a $d^{10}$ configuration), and Te is a metalloid (Te$^{4+}$ possesses a 5$s^2$ lone pair). The calculated $\tau$ values for these compounds, obtained using Shannon ionic radii~\cite{https://doi.org/10.1107/S0567739476001551}, are listed in Table~\textcolor{magenta}{S1} of the Supporting Information. With the exception of Te-based compounds, all systems exhibit $\tau$ values close to 1.0 and are therefore expected to crystallize in the cubic structure at room temperature.

As shown in Fig.~\ref{fig:st}, the Cs and $B$-site cations in cubic Cs$_2BX_6$ are 12-fold and 6-fold coordinated by the $X$-site anions, respectively. The Wyckoff positions of the Cs, $B$, and $X$ sites are 8$c$ (1/4, 1/4, 1/4), 4$b$ (1/2, 1/2, 1/2), and 24$e$ ($x$, 0, 0), with corresponding site symmetries $T_d$, $O_h$, and $C_{4v}$, respectively. As summarized in Table~\textcolor{magenta}{S1} of the Supporting Information, our calculated lattice constants ($a$) at 0~K are slightly smaller than the experimental values measured at 300~K, in good agreement when thermal expansion is taken into account. This agreement indicates the high reliability of our calculations and is consistent with our previous findings for binary semiconductors, where PBEsol was shown to accurately predict lattice constants when temperature effects are considered~\cite{PhysRevB.108.024306}. As $X$ changes from Cl to Br and then to I, the lattice constant $a$ increases slightly for compounds with the same $B$-site cation, reflecting the increasing ionic radius of the halide anion. Both experimental measurements and DFT calculations show that Cs$_2$Pd$X_6$ exhibits the smallest lattice constant among compounds with the same $X$ anion, whereas Cs$_2$Zr$X_6$ has the largest.

As discussed above, only the $X$-site Wyckoff position (24$e$) contains a free internal parameter, $x$. In the ideal structure ($x = 0.25$), the lattice constant $a$ equals $2\sqrt{2}$ times the $A$--$X$ bond length or four times the $B$--$X$ bond length. In practice, a mismatch between the ionic radii of the $A$- and $B$-site cations leads to deviations from this ideal value, resulting in a bending of the $A$--$X$--$A$ bond angle away from 180$^{\circ}$. Consequently, the $A$--$X$--$A$ angle serves as a direct measure of internal strain in these compounds and provides a more physically meaningful descriptor than the tolerance factor, which is based on empirical ionic radii. As shown in Table~\textcolor{magenta}{S1} of the Supporting Information, the deviation of the $A$--$X$--$A$ angle from 180$^{\circ}$ decreases systematically from Cs$_2B$Cl$_6$ to Cs$_2B$Br$_6$ and further to Cs$_2B$I$_6$. Among different $B$-site cations, Cs$_2$Pd$X_6$ and Cs$_2$Pt$X_6$ exhibit the largest deviations, whereas Cs$_2$Te$X_6$ shows the smallest deviation.

\begin{figure}[tph!]
	\includegraphics[clip,width=1.0\linewidth]{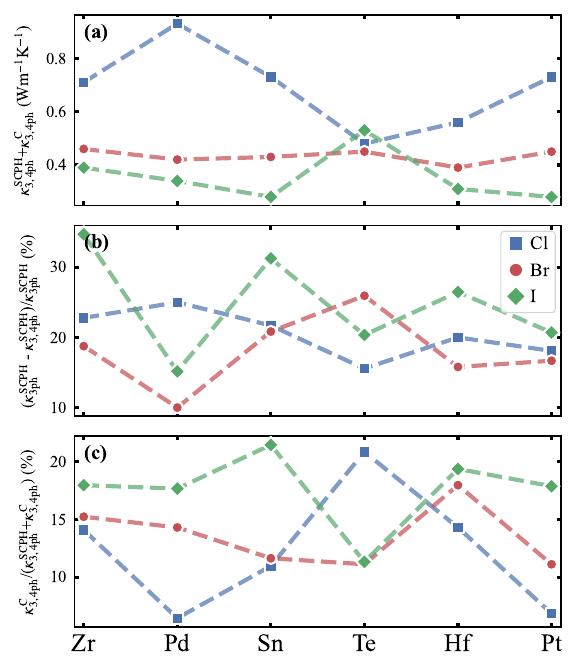}
	\caption{(a) Calculated lattice thermal conductivity $\kappa_{\mathrm{L}}$ at 300~K, including three- and four-phonon scattering processes ($\kappa_{\mathrm{3,4ph}}^{\mathrm{SCPH}}$) and the coherent contribution ($\kappa_{\mathrm{3,4ph}}^{\mathrm{C}}$), obtained using renormalized second-order force constants within the SCPH framework. (b) Contribution of three-phonon scattering processes ($\kappa_{\mathrm{3ph}}^{\mathrm{SCPH}}$) to $\kappa_{\mathrm{L}}$. (c) Contribution of the coherent term ($\kappa_{\mathrm{3,4ph}}^{\mathrm{C}}$) to the total $\kappa_{\mathrm{L}}$.}
	\label{fig:kappa}
\end{figure}

\begin{figure*}
	\centering
	\includegraphics[width=1.0\linewidth]{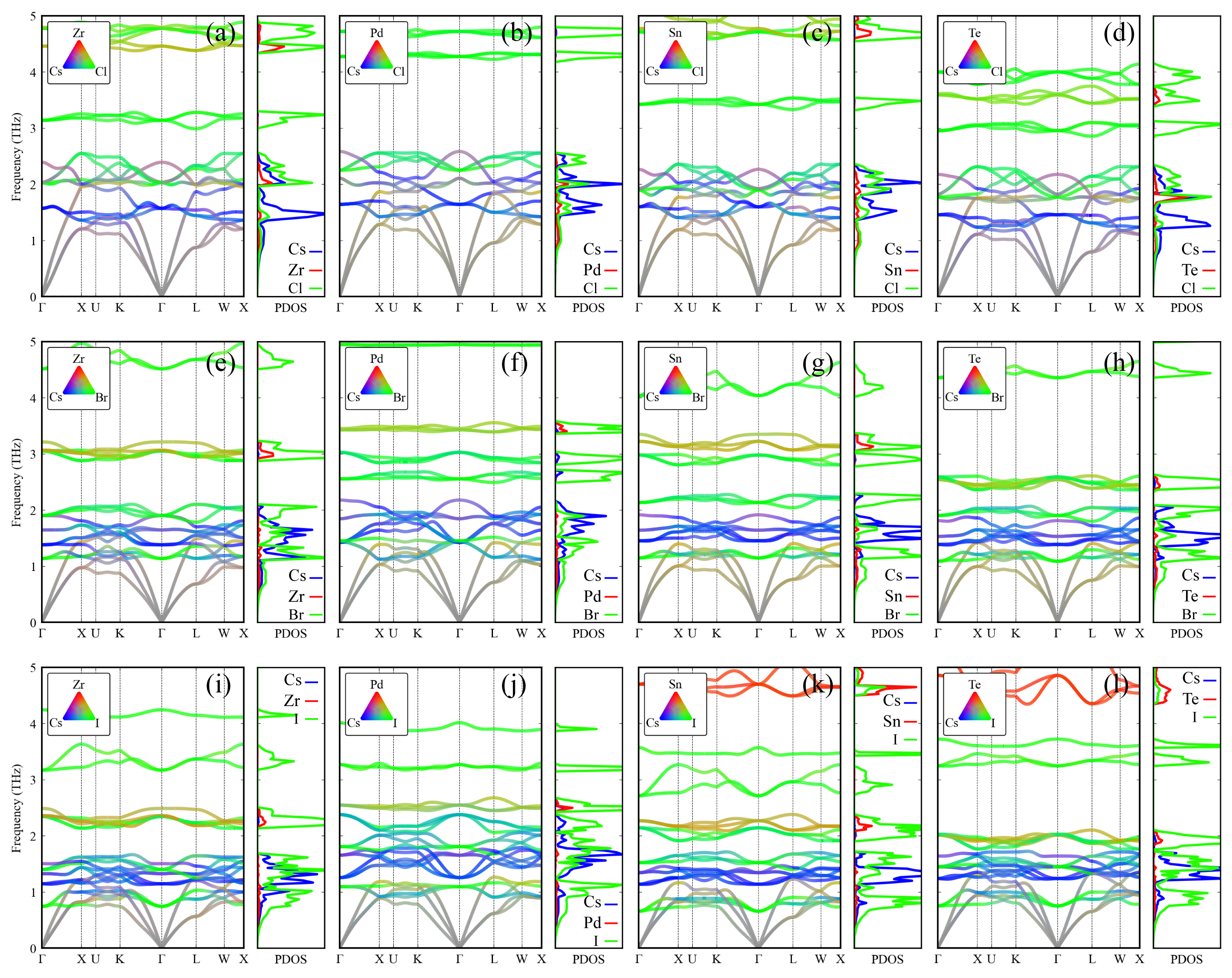}
	\caption{Phonon band structures and phonon density of states (PhDOS) of Cs$_2BX_6$ ($B$ = Zr, Pd, Sn, and Te; $X$ = Cl, Br, and I) calculated at 300~K. Panels (a)–(d) show the phonon band structures and PhDOS of Cs$_2B$Cl$_6$ with $B$ = Zr, Sn, Pd, and Te, respectively. Panels (e)–(h) present the corresponding results for Cs$_2B$Br$_6$, and panels (i)–(l) for Cs$_2B$I$_6$, with the same ordering of $B$ cations.}
	\label{fig:band-dos}
\end{figure*}

\begin{figure*}[tph!]
	\includegraphics[clip,width=1.0\linewidth]{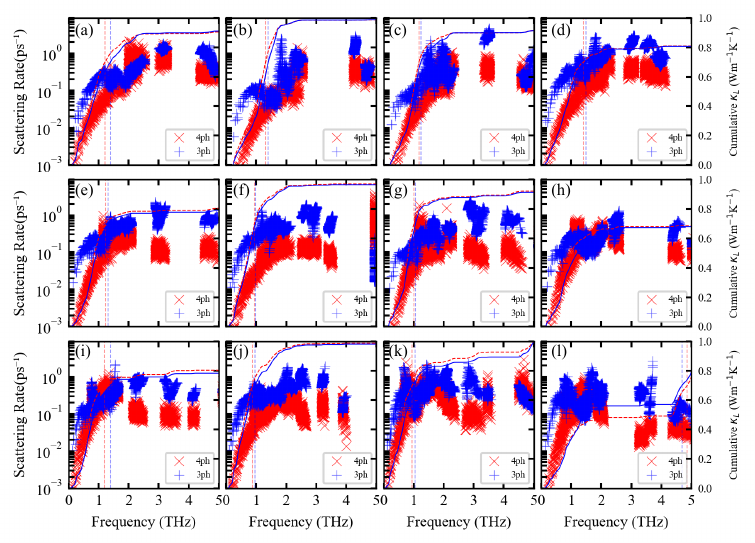}
	\caption{Three- and four-phonon scattering rates of Cs$_2BX_6$ ($B$ = Zr, Pd, Sn, and Te; $X$ = Cl, Br, and I) calculated at 300~K. Panels (a)–(d) show the phonon scattering rates of Cs$_2B$Cl$_6$ with $B$ = Zr, Sn, Pd, and Te, respectively. Panels (e)–(h) present the corresponding results for Cs$_2B$Br$_6$, and panels (i)–(l) for Cs$_2B$I$_6$, with the same ordering of $B$ cations.}
	\label{fig:scattering}
\end{figure*}

\subsection{Lattice thermal conductivity at room temperature}
Our calculated $\kappa_{\mathrm{L}}$ of Cs$_2BX_6$ ($B$ = Zr, Pd, Sn, Te, Hf, and Pt; $X$ = Cl, Br, and I) at 300~K are shown in Fig.~\ref{fig:kappa}. The results include $\kappa_{\mathrm{L}}$ evaluated at two advanced rungs of Jacob’s ladder~\cite{PhysRevX.10.041029}: (i) three-phonon scattering with renormalized second-order FCs obtained from the SCPH method ($\kappa_{\mathrm{3ph}}^{\mathrm{SCPH}}$), and (ii) combined three- and four-phonon scattering with renormalized second-order FCs ($\kappa_{\mathrm{3,4ph}}^{\mathrm{SCPH}}$). In addition, the coherent contributions to the thermal conductivity, computed by including three-phonon ($\kappa_{\mathrm{3ph}}^{\mathrm{C}}$) and three- and four-phonon ($\kappa_{\mathrm{3,4ph}}^{\mathrm{C}}$) interactions, are also presented. Our calculated $\kappa_{\mathrm{L}}$ of Cs$_2$SnI$_6$ at room temperature (0.28~W\,m$^{-1}$\,K$^{-1}$) is in excellent agreement with the experimental value of 0.29~W\,m$^{-1}$\,K$^{-1}$~\cite{doi:10.1021/acs.chemmater.2c00084}. Similarly, the calculated $\kappa_{\mathrm{L}}$ of Cs$_2$SnBr$_6$ (0.47~W\,m$^{-1}$\,K$^{-1}$) is close to a previous theoretical result of 0.43~W\,m$^{-1}$\,K$^{-1}$ obtained using a comparable methodology~\cite{PhysRevB.109.054305}.

As shown in Fig.~\ref{fig:kappa}(a), all compounds exhibit intrinsically low $\kappa_{\mathrm{L}}$ ($< 1.0$~W\,m$^{-1}$\,K$^{-1}$), with the highest and lowest values observed for Cs$_2$PdCl$_6$ and Cs$_2$SnI$_6$, respectively. For most compounds, $\kappa_{\mathrm{L}}$ decreases systematically as $X$ changes from Cl to Br and then to I, which can be attributed to the increasing atomic mass of the halide and the concomitant weakening of the $A$--$X$ and $B$--$X$ ionic bonds. An exception to this trend is found in the Cs$_2$Te$X_6$ series, where Cs$_2$TeI$_6$ exhibits a slightly higher $\kappa_{\mathrm{L}}$ than Cs$_2$TeCl$_6$ and Cs$_2$TeBr$_6$. This behavior originates from the larger particle-like contributions $\kappa_{\mathrm{3ph}}^{\mathrm{P}}$ and $\kappa_{\mathrm{3,4ph}}^{\mathrm{P}}$ in Cs$_2$TeI$_6$, as summarized in Table~\textcolor{magenta}{S1}. Although the average sound velocity $\nu_{\mathrm{g}}$ decreases monotonically from Cl to I, this anomalous evolution of $\kappa_{\mathrm{L}}$ suggests stronger phonon--phonon interactions in Cs$_2$TeCl$_6$ and Cs$_2$TeBr$_6$, as discussed in the next section.

Overall, the $\kappa_{\mathrm{L}}$ of Cs$_2B$Cl$_6$ compounds exhibit larger variations with respect to the choice of $B$ than those of the Cs$_2B$Br$_6$ and Cs$_2B$I$_6$ counterparts. In contrast, for Cs$_2B$Br$_6$ and Cs$_2B$I$_6$ compounds, $\kappa_{\mathrm{L}}$ depends only weakly on the $B$-site cation, with the notable exception of the anomalously high $\kappa_{\mathrm{L}}$ in Cs$_2$TeI$_6$. These trends indicate that the halide species $X$ plays a dominant role in determining $\kappa_{\mathrm{L}}$ in Cs$_2B$Cl$_6$ compounds, whereas its influence is less pronounced in Cs$_2B$Br$_6$ and Cs$_2B$I$_6$ systems.

The relative importance of four-phonon scattering is quantified by the ratio
$(\kappa_{\mathrm{3ph}}^{\mathrm{SCPH}} - \kappa_{\mathrm{3,4ph}}^{\mathrm{SCPH}})/\kappa_{\mathrm{3ph}}^{\mathrm{SCPH}}$.
As shown in Fig.~\ref{fig:kappa}(b), four-phonon processes reduce the $\kappa_{\mathrm{L}}$ by approximately 20--25\% in most compounds, comparable to previous observations in zinc-blende semiconductors~\cite{PhysRevB.110.035205}. The smallest and largest reductions are found in Cs$_2$PdBr$_6$ (10\%) and Cs$_2$ZrCl$_6$ (35\%), respectively. All compounds exhibit pronounced variations in this ratio with respect to both $B$ and $X$. Notably, Cs$_2B$I$_6$ compounds generally display the largest four-phonon contribution for all $B$ except Pd and Te, whereas Cs$_2B$Br$_6$ compounds show the smallest contribution, except for the Cs$_2$Te$X_6$ compounds. This behavior is likely associated with the relatively small values of $\kappa_{\mathrm{3ph}}^{\mathrm{SCPH}}$ in the Cs$_2B$I$_6$ compounds.

Although all Cs$_2BX_6$ compounds studied here possess low $\kappa_{\mathrm{L}}$, the coherent contribution $\kappa_{\mathrm{3,4ph}}^{\mathrm{C}}$ remains relatively small at room temperature. For most systems, the ratio $\kappa_{\mathrm{3,4ph}}^{\mathrm{C}}/(\kappa_{\mathrm{3,4ph}}^{\mathrm{P}} + \kappa_{\mathrm{3,4ph}}^{\mathrm{C}})$ is approximately 15\%, with the largest value observed for Cs$_2$SnI$_6$ (21\%). In general, Cs$_2B$I$_6$ compounds exhibit larger $\kappa_{\mathrm{3,4ph}}^{\mathrm{C}}$ fractions than Cs$_2B$Cl$_6$ and Cs$_2B$Br$_6$ compounds, except for Cs$_2$Te$X_6$ compounds, where the ratio decreases from Cl to Br and then increases from Br to I. The magnitude of the $\kappa_{\mathrm{3,4ph}}^{\mathrm{C}}$ contribution is closely related to lattice anharmonicity. Cs$_2$PdCl$_6$ and Cs$_2$PtCl$_6$ display the smallest coherent contributions (6\% and 7\%, respectively). Overall, the $\kappa_{\mathrm{3,4ph}}^{\mathrm{C}}$ contributions in these materials are significantly smaller than those reported for several ultralow-$\kappa_{\mathrm{L}}$ compounds, such as NaAg$_3$Se$_2$ ($\sim$40\%)~\cite{https://doi.org/10.1002/advs.202417292}, Cs$_3$Bi$_2$Br$_9$ ($\sim$50\%)~\cite{https://doi.org/10.1002/adfm.202411152}, and CsCu$_2$I$_3$ ($>$50\%)~\cite{shen2025accelerated}. This comparison indicates that the lattice anharmonicity in Cs$_2BX_6$ is moderate and weaker than that typically observed in other halide perovskites~\cite{PhysRevLett.125.045701}.

\subsection{Phonon dispersion and phonon-phonon scattering at room temperature}
Because $\kappa_{\mathrm{L}}$ is strongly correlated with phonon dispersion, we first examine the phonon spectra of these compounds to elucidate the microscopic origin of their low $\kappa_{\mathrm{L}}$. The phonon dispersions and phonon densities of states (PhDOS) of Cs$_2BX_6$ ($B$ = Zr, Pd, Sn, and Te; $X$ = Cl, Br, and I) calculated at 300~K within the SCPH framework are shown in Fig.~\ref{fig:band-dos}. Owing to their close similarity to Cs$_2$Zr$X_6$ and Cs$_2$Pd$X_6$, the corresponding results for Cs$_2$Hf$X_6$ and Cs$_2$Pt$X_6$ are presented in Fig.~\textcolor{magenta}{S1} of the Supporting Information. Phonon dispersions at other temperatures are also provided in Fig.~\textcolor{magenta}{S2}.

For all compounds, the lowest optical phonon branches are relatively flat and exhibit low frequencies, lying below the highest acoustic phonon modes. Specifically, the frequencies of these modes at the $\Gamma$ point are approximately 1.5~THz for Cs$_2B$Cl$_6$, 1.2~THz for Cs$_2B$Br$_6$, and 0.8~THz for Cs$_2B$I$_6$. This systematic reduction reflects both the increasing atomic mass and the decreasing electronegativity of the halide anion from Cl to I. In Cs$_2B$Cl$_6$, the lowest optical modes are dominated by vibrations of the Cs atoms, whereas in Cs$_2B$Br$_6$ and Cs$_2B$I$_6$ they are primarily associated with the heavier $X$ atoms. Interestingly, the frequencies of the lowest optical modes exhibit a noticeable discontinuity at $B$ = Pd when moving along the sequence of increasing atomic number from Zr to Te. The maximum frequencies of the longitudinal acoustic (LA) phonons at the Brillouin-zone boundaries (X, K, and L points) are below 2~THz for Cs$_2B$Cl$_6$, and are further reduced in Cs$_2B$Br$_6$ and Cs$_2B$I$_6$. This behavior is consistent with the low average sound velocities $\nu_{\mathrm{g}}$ listed in Table~\textcolor{magenta}{S1} of the Supporting Information. The presence of low-lying optical phonons enhances optical--acoustic phonon scattering, which plays a key role in suppressing $\kappa_{\mathrm{L}}$.

The calculated three- and four-phonon scattering rates are shown in Fig.~\ref{fig:scattering}. Because the phonon scattering characteristics of Cs$_2$Hf$X_6$ and Cs$_2$Pt$X_6$ closely resemble those of Cs$_2$Zr$X_6$ and Cs$_2$Pd$X_6$, respectively, their results are presented in Fig.~\textcolor{magenta}{S3} of the Supporting Information. For all compounds, the three-phonon scattering rates in the low-frequency range (0--1.5~THz), which contributes more than 70\% of the total particle-like $\kappa_{\mathrm{L}}$, are on the order of $10^{12}$~s$^{-1}$. These values are substantially smaller than those in archetypal systems with strong anharmonicity, such as AgCl~\cite{PhysRevX.10.041029}, R-Heusler compounds~\cite{PhysRevB.110.035205}, NaAg$_3$S$_2$~\cite{https://doi.org/10.1002/advs.202417292}, and skutterudite YbFe$_4$Sb$_{12}$~\cite{PhysRevB.91.144304}. Moreover, in the low-frequency region around $\sim$1.2~THz, where the cumulative $\kappa_{\mathrm{L}}$ reaches approximately 80\%, the four-phonon scattering rates are generally much smaller than the corresponding three-phonon rates. Overall, nearly all compounds exhibit progressively stronger phonon--phonon scattering when $X$ changes from Cl to Br and then to I. This trend is consistent with the decreasing electronegativity, weakening chemical bonding, and increasing anharmonicity associated with heavier halide anions, which enhance both three- and four-phonon interactions. As shown in Fig.~\textcolor{magenta}{S4}, the relatively large four-phonon scattering rates in Cs$_2B$I$_6$ compounds can be attributed to their large weighted phase space~\cite{PhysRevB.91.144304}, arising from flat low-frequency optical branches dominated by I-atom vibrations. As illustrated in Fig.~\ref{fig:scattering}, Cs$_2$SnI$_6$ exhibits the strongest three- and four-phonon scattering rates in the frequency range of 0.5--1.0~THz, corresponding to a large phase space (Fig.~\textcolor{magenta}{S4}). Because this frequency range accounts for approximately 80\% of the total heat transport, the enhanced scattering leads to the lowest $\kappa_{\mathrm{L}}$ among the compounds studied. Similarly, the lower $\kappa_{\mathrm{L}}$ of Cs$_2$TeCl$_6$ compared with Cs$_2$TeI$_6$ originates from the relatively strong three-phonon scattering near 2.0~THz.

\begin{figure*}[tph!]
	\includegraphics[clip,width=1.0\linewidth]{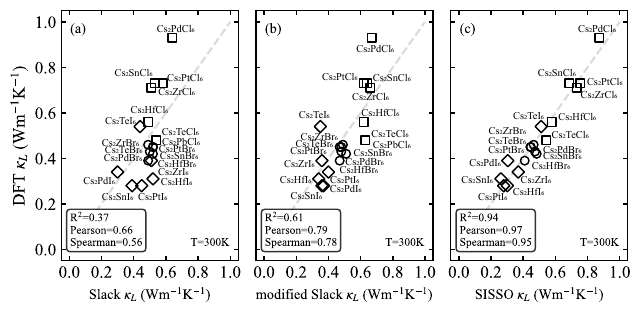}
	\caption{Comparison between the DFT-calculated $\kappa_{\mathrm{L}}$ and the predicted $\kappa_{\mathrm{L}}$ of Cs$_2BX_6$ ($B$ = Zr, Pd, Sn, Te, Hf, and Pt; $X$ = Cl, Br, and I) obtained using different models. Panels (a)–(c) correspond to the Slack, modified Slack, and SISSO-constructed models, respectively.}
	\label{fig:speed_of_sound_vs_kappa}
\end{figure*}

\subsection{Machine learning modes of $\kappa_{\mathrm{L}}$ at room temperature}
With 18 our calculated $\kappa_{\mathrm{L}}$ using state-of-the-art first-principles approaches, we first assess the semi-empirical model proposed by Slack~\cite{SLACK1973321} and the modified model introduced by Toberer \emph{et al.}~\cite{C1JM11754H}. We then construct a new predictive model assisted by a machine-learning approach.  
Within the Slack model, only acoustic phonon scattering is considered, and $\kappa_{\mathrm{L}}$ is expressed as
\begin{equation}
	\kappa_{\mathrm{L}} = A_0\frac{\overline{M}\nu_g^{3}\delta n^{1/3}}{\gamma^{2}T},
\end{equation}
\noindent where $\overline{M}$, $\delta^3$, $n$, $\gamma$, and $T$ denote the average atomic mass, volume per atom, number of atoms in the primitive cell, Gr{\"u}neisen parameter, and temperature, respectively.  
The performance of the Slack model, evaluated using the parameters listed in Table~\textcolor{magenta}{S1}, is shown in Fig.~\ref{fig:speed_of_sound_vs_kappa}(a). The resulting $R^2$, Pearson, and Spearman coefficients are 0.39, 0.66, and 0.56, respectively. The optimized fitting parameter $A_0$ is $1.82\times10^{-6}$, which differs substantially from the original value of $3.04\times10^{-8}$.

To account for phonon scattering from both acoustic ($\kappa_{\mathrm{L}}^{\mathrm{a}}$) and optical branches ($\kappa_{\mathrm{L}}^{\mathrm{o}}$), Toberer \emph{et al.}~\cite{C1JM11754H} proposed an improved model containing two fitting parameters, $A_0$ and $A_1$,
\begin{equation}
	\kappa_{\mathrm{L}} = \kappa_{\mathrm{L}}^{\mathrm{a}} + \kappa_{\mathrm{L}}^{\mathrm{o}}
	= A_0\frac{\overline{M}\nu_{\mathrm{g}}^{3}}{V^{2/3}n^{1/3}}
	+ A_1\frac{\nu_{\mathrm{g}}}{V^{2/3}}\left(1-\frac{1}{n^{2/3}}\right),
\end{equation}
\noindent where $\overline{M}$, $V$, $n$, and $\nu_{\mathrm{g}}$ represent the average atomic mass, average volume per atom, number of atoms in the primitive cell, and sound velocity, respectively.  
As shown in Fig.~\ref{fig:speed_of_sound_vs_kappa}(b), this model yields improved agreement, with $R^2$, Pearson, and Spearman coefficients of 0.61, 0.79, and 0.78, respectively. The fitted parameters ($A_0=-4.95\times10^{-4}$, $A_1=2.15\times10^{-22}$) differ from the original values ($A_0=2.7\times10^{-4}$, $A_1=1.5\times10^{-23}$) reported in Ref.~\cite{C1JM11754H}. Overall, the modified Slack model provides significantly higher accuracy than the original Slack model for the compounds considered here.

\begin{figure*}[tph!]
	\includegraphics[clip,width=0.8\linewidth]{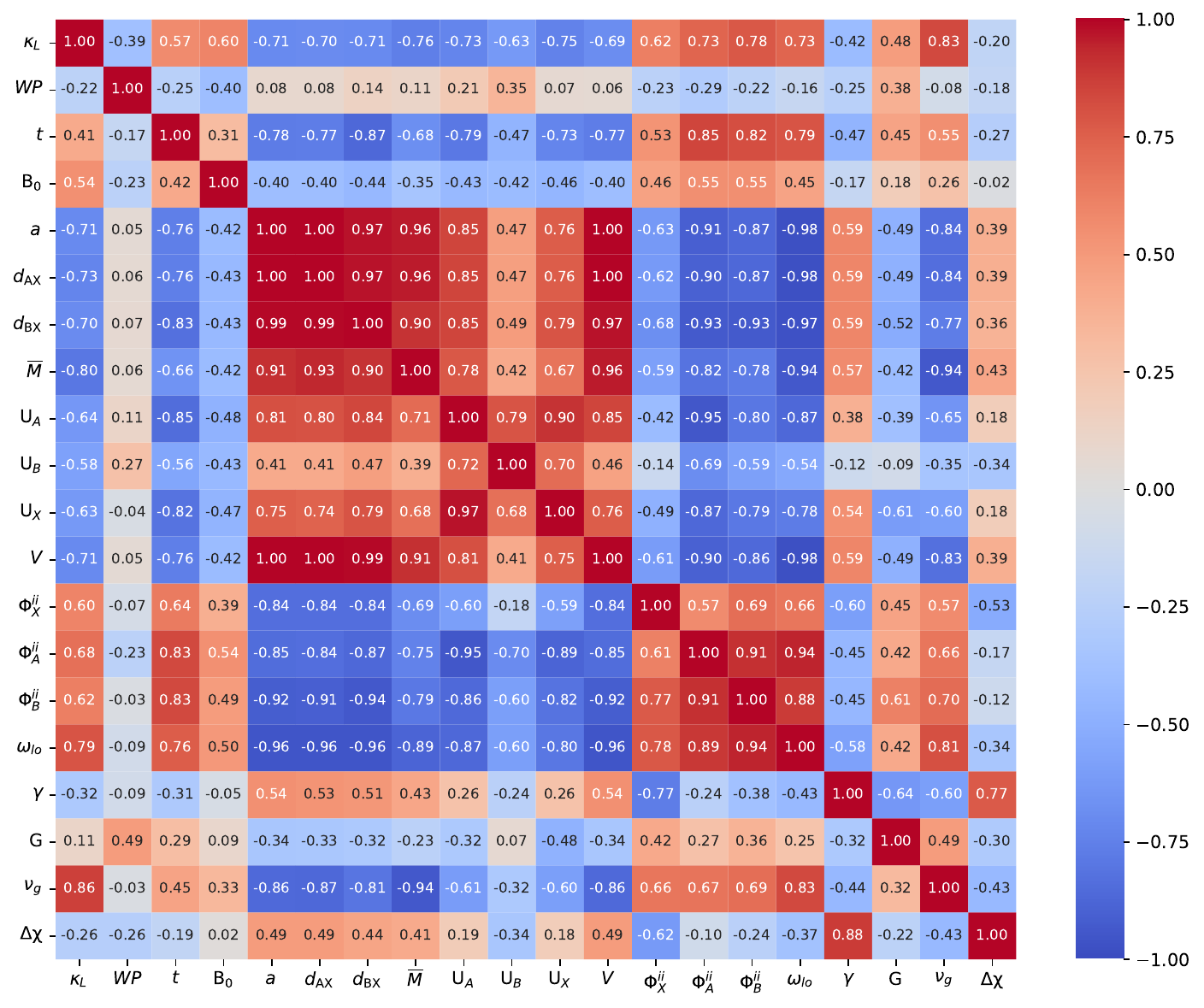}
	\caption{Correlation matrix between the features used in this work and the $\kappa_{\mathrm{L}}$. The upper (lower) triangle shows the Spearman (Pearson) correlation coefficients. Here, WP denotes the Wyckoff position of $X$; $\tau$ is the tolerance factor; $B_0$ is the bulk modulus; $a$ is the lattice constant; $d_{AX}$ ($d_{BX}$) is the distance between the $A$ ($B$) and $X$ atoms; $\overline{M}$ is the average atomic mass; $U_A$, $U_B$, and $U_X$ are the mean-square displacements of the $A$, $B$, and $X$ atoms, respectively; $\Phi_A^{ii}$, $\Phi_B^{ii}$, and $\Phi_X^{ii}$ are the root-mean-square on-site force constants of the $A$, $B$, and $X$ atoms, respectively; $\omega_{\mathrm{lo}}$ is the lowest optical phonon frequency; $\gamma$ is the Gr\"uneisen parameter; and $\Delta \chi$ is the electronegativity difference between the $B$ and $X$ atoms.}
	\label{fig:feature}
\end{figure*}

To develop a machine-learning model for $\kappa_{\mathrm{L}}$ based on this relatively small dataset, we selected descriptors from Table~\textcolor{magenta}{S1} of the Supplemental Material. We first filtered out mutually correlated features using Pearson and Spearman correlation coefficients, see Fig.~\ref{fig:feature}. When two features exhibited a correlation coefficient larger than 0.8, only the feature more strongly correlated with $\kappa_{\mathrm{L}}$ was retained. Ultimately, four descriptors were selected for model construction. The machine-learning model for $\kappa_{\mathrm{L}}$ was then constructed using the Sure Independence Screening and Sparsifying Operator (SISSO)~\cite{PhysRevMaterials.2.083802}, yielding the following expression:
\begin{equation}
	\kappa_{\mathrm{L}} = A_0\frac{B_0^{2}\nu_{\mathrm{g}}}{U_{B}^{1/3}} + A_1,
\end{equation}
\noindent where $A_0$ and $A_1$ are fitting parameters with optimized values $A_0$ = 1.66 $\times$ 10$^{-6}$ and $A_1$ = -0.60, respectively. Here, $\nu_{\mathrm{g}}$, $B_0$, and $U_B$ denote the sound velocity, bulk modulus, and mean-square displacement (MSD) of the $B$-site atom. As illustrated in Fig.~\ref{fig:speed_of_sound_vs_kappa}(c), this SISSO model significantly outperforms both the Slack and modified Slack models, achieving an $R^2$ of 0.96 and Pearson and Spearman coefficients of 0.98 and 0.96, respectively. Although $B_0$ and $U_B$ individually show relatively weak correlations with $\kappa_{\mathrm{L}}$, their combination with $\nu_{\mathrm{g}}$ maximizes the predictive power, highlighting the crucial role of chemical bond strength in determining $\kappa_{\mathrm{L}}$.

\begin{figure*}[tph!]
	\includegraphics[clip,width=1.0\linewidth]{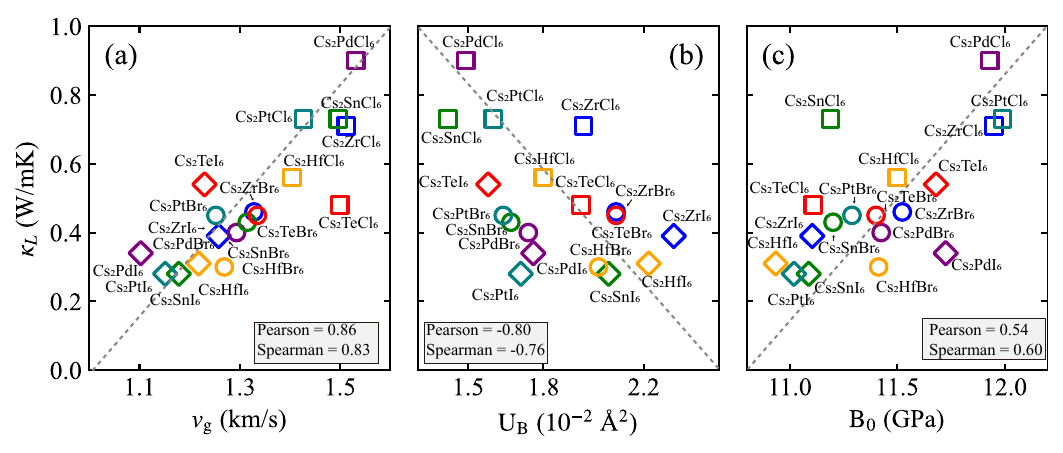}
	\caption{Dependence of the total lattice thermal conductivity $\kappa_{\mathrm{L}}$ on (a) the phonon group velocity $\nu_{\mathrm{g}}$, (b) the mean-square displacement of the $B$-site atom ($U_B$), and (c) the bulk modulus ($B_0$).} 
	\label{fig:kappa-mass-speed}
\end{figure*}

Notably, our model predicts $\kappa_{\mathrm{L}} \propto \nu_{\mathrm{g}}$, in contrast to the $\nu_{\mathrm{g}}^{3}$ dependence in the Slack and modified Slack models. This difference is likely due to the relatively narrow range of $\nu_{\mathrm{g}}$ (1100--1600~m/s) spanned by these compounds. The correlation between $\kappa_{\mathrm{L}}$ and $\nu_{\mathrm{g}}$ is shown in Fig.~\ref{fig:kappa-mass-speed}(a). Cs$_2$TeCl$_6$ exhibits the largest deviation from the linear trend, indicating a substantially lower $\kappa_{\mathrm{L}}$ than expected from its $\nu_{\mathrm{g}}$. This observation is consistent with previous results showing anomalously low $\kappa_{\mathrm{L}}$ in Cs$_2$TeCl$_6$ within the Cs$_2B$Cl$_6$ series (see Fig.~\ref{fig:kappa}), which originates from strong three-phonon scattering in the low-frequency region (Fig.~\ref{fig:scattering}).

\begin{figure*}[tph!]
	\includegraphics[clip,width=1.0\linewidth]{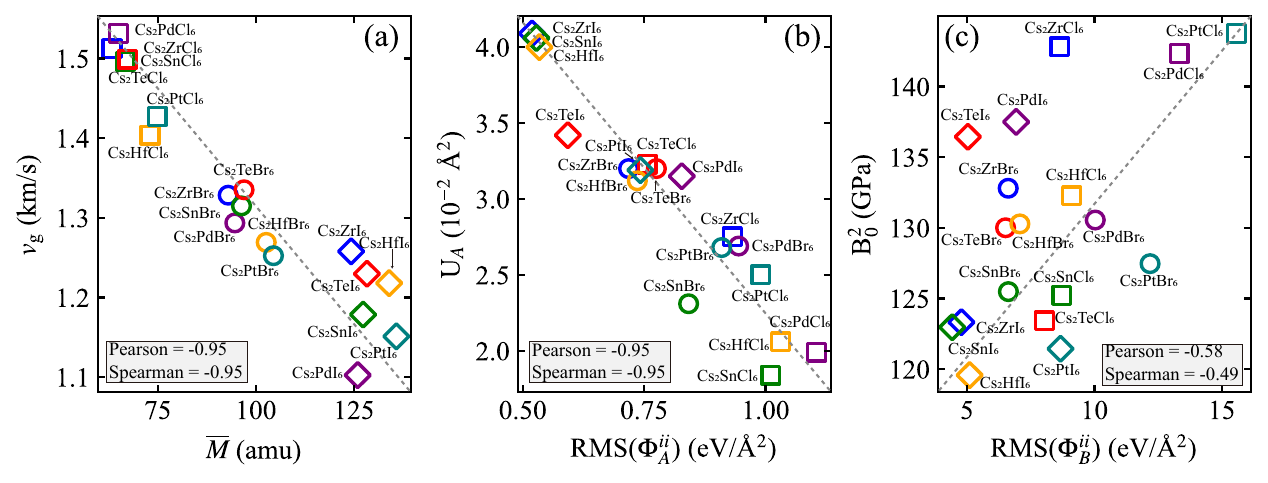}
	\caption{Dependence of (a) the phonon group velocity $\nu_{\mathrm{g}}$ on the average atomic mass $\overline{M}$, (b) the mean-square displacement of the $A$-site atom ($U_A$) on the root-mean-square on-site force constant $\mathrm{RMS}(\Phi_A^{ii})$, and (c) the atomic mass $M$ on the root-mean-square on-site force constant $\mathrm{RMS}(\Phi_B^{ii})$.}
	\label{fig:kappa-xsite-rms}
\end{figure*}

The average atomic mass $\overline{M}$ does not explicitly appear in our SISSO model, even though $\kappa_{\mathrm{L}}$ is strongly correlated with $\overline{M}$ (Spearman: $-0.76$, Pearson: $-0.80$; Fig.~\ref{fig:feature}). This can be understood from the strong correlation between $\nu_{\mathrm{g}}$ and $\overline{M}$ (Spearman: $-0.95$, Pearson: $-0.95$), as shown in Fig.~\ref{fig:kappa-xsite-rms}(a). Such behavior is expected since $\nu_{\mathrm{g}} \propto \sqrt{K/\overline{M}}$, and the high correlation coefficients indicate similar bond strengths across this family of compounds. The relationship between $\kappa_{\mathrm{L}}$ and $U_B$ is shown in Fig.~\ref{fig:kappa-mass-speed}(b), where larger $U_B$ generally corresponds to lower $\kappa_{\mathrm{L}}$. A larger $U_B$ reflects weaker chemical bonding between the $B$-site cation and surrounding anions and is often associated with pronounced rattling modes. Although $U_A$ is typically larger than $U_B$ due to weaker $A$--$X$ bonding (Table~\textcolor{magenta}{S1}), $U_A$, $U_B$, and $U_X$ exhibit similar correlations with $\kappa_{\mathrm{L}}$ (Pearson: $-0.64$, $-0.58$, and $-0.63$, respectively; Fig.~\ref{fig:feature}). Among them, $U_B$ alone shows the weakest correlation with $\kappa_{\mathrm{L}}$, yet its combination with $B_0$ yields the highest predictive accuracy. Cs$_2$PtI$_6$ shows the largest deviation from the linear $\kappa_{\mathrm{L}}$--$U_B$ trend, with a smaller $\kappa_{\mathrm{L}}$ than expected from its $U_B$ value. The large magnitude of $U_A$ can be traced to the bonding between the $A$-site cation and its surrounding ions, which is characterized by the on-site second-order FCs $\Phi_{A}^{\mathrm{ii}}$. As shown in Fig.~\ref{fig:kappa-xsite-rms}(b), $U_A$ is strongly correlated with the root-mean-square (RMS) of FCs, $\Phi_{A}^{\mathrm{ii}}$~\cite{Qin2018}. The $U_A$, $U_B$, and $U_X$ are also mutually correlated, with correlation coefficients of approximately 0.70.

Additional features with strong correlations to $\kappa_{\mathrm{L}}$ include the lowest optical phonon frequency at the $\Gamma$ point ($\omega_{\mathrm{lo}}$), which correlates strongly with both $\nu_{\mathrm{g}}$ and $\Phi_{B}^{\mathrm{ii}}$, and the $A$--$X$ and $B$--$X$ bond lengths ($d_{A\text{-}X}$ and $d_{B\text{-}X}$; Spearman: $-0.71$, Pearson: $-0.70$). However, these descriptors were not selected by SISSO because of their strong correlations with $\nu_{\mathrm{g}}$ ($\sim$0.85) and $\overline{M}$ ($\sim$0.9). Moreover, $d_{A\text{-}X}$ and $d_{B\text{-}X}$ are strongly correlated with the lattice constant $a$, since deviations of the internal coordinate $x$ from its ideal value (0.25) are negligible (Table~\textcolor{magenta}{S1}).

Although $B_0$ alone correlates weakly with $\kappa_{\mathrm{L}}$ than $\nu_{\mathrm{g}}$ and $U_B$, SISSO identifies $B_0^2$ as a critical descriptor. Correlation analysis shows that $B_0$ is most strongly correlated with $\Phi_{A}^{\mathrm{ii}}$ (Pearson: 0.54, Spearman: 0.55) and $\Phi_{B}^{\mathrm{ii}}$ (Pearson: 0.49, Spearman: 0.55). Given that $\Phi_{B}^{\mathrm{ii}}$ is much larger in magnitude than $\Phi_{A}^{\mathrm{ii}}$, Fig.~\ref{fig:kappa-xsite-rms} presents the correlation between $B_0$ and $\Phi_{B}^{\mathrm{ii}}$. Cs$_2$TeI$_6$, Cs$_2$PdI$_6$, and Cs$_2$ZrCl$_6$ show notable deviations from the linear trend, likely reflecting anomalous behavior in the $B_0$ of Cs$_2$Pd$X_6$ and Cs$_2$Te$X_6$. For example, Cs$_2$PdI$_6$ exhibits a larger $B_0$ than Cs$_2$PdBr$_6$, despite a larger electronegativity difference between Pd and Br than between Pd and I. Similarly, $B_0$ in Cs$_2$Te$X_6$ increases unexpectedly as the electronegativity difference decreases from Cl to Br to I (Table~\textcolor{magenta}{S1}).

In summary, considering the similar magnitudes and correlations of the MSDs at the $A$, $B$, and $X$ sites (Fig.~\textcolor{magenta}{S5}), together with the weak correlation between $\gamma$ and $\kappa_{\mathrm{L}}$ (Pearson coefficient of 0.14), we conclude that the intrinsically low $\kappa_{\mathrm{L}}$ observed in these compounds primarily originates from their low $\nu_{\mathrm{g}}$.

\subsection{Chemical bonding trends}
In this section, we elucidate the chemical origin underlying the observations discussed above. Crystal orbital Hamilton population (COHP) analysis is widely employed to quantify chemical bonding characteristics. The $-$COHP curves and corresponding bond indices (BIs) for the present compounds are shown in Fig.~\textcolor{magenta}{S6} and summarized in Table~\textcolor{magenta}{S1} of the Supplemental Material, respectively. Similar to conventional perovskites, the strongest chemical bonding occurs between the $B$-site cation and the $X$-site anion, as evidenced by the largest integrated COHP ($-$iCOHP) values. This interaction is followed in strength by the Cs--$X$ and Cs--$B$ bonds, as well as by the on-site FCs. In addition, the $B$--$X$ BI systematically decreases as the halogen is varied from Cl to Br to I.

When comparing different $B$-site cations, the $-$iCOHP values exhibit pronounced variations. Within an octahedral crystal field, the $d$ orbitals of the $B$-site cation split into $t_{2g}$ and $e_g$ manifolds. For Zr$^{4+}$, the $4d$ orbitals are empty; consequently, the antibonding states ($t_{2g}$ and $e_g$) formed by hybridization between Zr-$d$ and $X$-$p$ orbitals remain unoccupied. In contrast, Pd$^{4+}$ possesses six $d$ electrons, and the strong crystal-field splitting characteristic of $4d$ elements favors a low-spin configuration ($t_{2g}^{\uparrow\uparrow\uparrow\downarrow\downarrow\downarrow}e_g^{0}$). As a result, the antibonding $\pi^*$ states dominated by $t_{2g}$ orbitals are fully occupied, while the $\sigma^*$ antibonding states associated with $e_g$ orbitals remain empty~\cite{Woodward_Karen_Evans_Vogt_2021}. This occupation of antibonding states leads to smaller $-$iCOHP values in Cs$_2$Pd$X_6$ compounds compared with their Cs$_2$Zr$X_6$ counterparts.

In Cs$_2$Sn$X_6$ compounds, the coupling between Sn-$d$ and $X$-$p$ orbitals is very weak~\cite{https://doi.org/10.1002/adfm.202108532}, due to the large energy difference between Sn-$d$ and $X$-$p$ orbitals. Consequently, Sn$^{4+}$ systems behave similarly to compounds dominated by $s$ and $p$ electrons, with only a small number of occupied antibonding states below the Fermi level. This results in comparatively large $-$iCOHP values, as shown in Fig.~\textcolor{magenta}{S6}. The valence-band maximum, characterized by the irreducible representation $T_{1u}$, is primarily derived from $X$-$p$ orbitals. A similar argument applies to Cs$_2$Te$X_6$ compounds: the $p$--$d$ coupling between Te-$d$ and $X$-$p$ orbitals is negligible because of their large energy separation. In Cs$_2$Te$X_6$, the Te$^{4+}$ cation hosts $5s^2$ lone-pair electrons, which give rise to strongly anti-bonding states below the Fermi level~\cite{C1CS15098G}, thereby reducing the $-$iCOHP values. As a consequence, Sn--$X$ bonds exhibit larger $-$iCOHP values than those in Pd- and Te-based compounds. Finally, Hf- and Pt-based compounds closely resemble the Zr- and Pd-based systems, respectively, reflecting their similar electronic configurations.

\begin{figure}[tph!]
	\includegraphics[clip,width=1.0\linewidth]{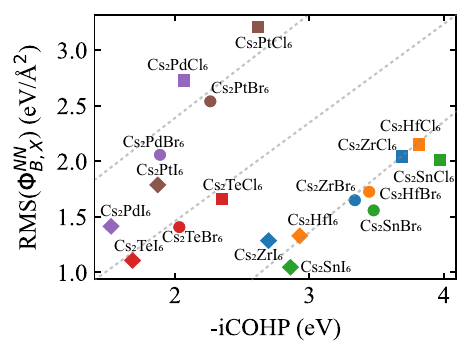}
	\caption{Correlation between the root-mean-square on-site force constant $\mathrm{RMS}(\Phi_{B}^{ii})$ and the integrated crystal orbital Hamilton population (iCOHP) of the $B$--$X$ bond.}
	\label{fig:icohp-2ndrms}
\end{figure}

The correlation between $-$iCOHP and the RMS values of the second-order interatomic FCs between $B$ and $X$ ions, RMS($\Phi_{B\text{-}X}^{\mathrm{NN}}$), is shown in Fig.~\ref{fig:icohp-2ndrms}. For a fixed $B$-site cation, RMS($\Phi_{B\text{-}X}^{\mathrm{NN}}$) depends nearly linearly on $-$iCOHP, consistent with the interpretation that a larger $-$iCOHP corresponds to stronger $B$--$X$ bonding. However, distinct trends emerge for different $B$-site cations: Zr-, Sn-, and Hf-based compounds follow a similar trend, as do Pd- and Pt-based compounds, whereas Te-based compounds clearly deviate. These observations are fully consistent with the differences in electronic configurations of the $B$-site cations discussed above.

\subsection{Temperature dependence of lattice thermal conductivity}
\begin{figure}
	\centering
	\includegraphics[width=1.0\linewidth]{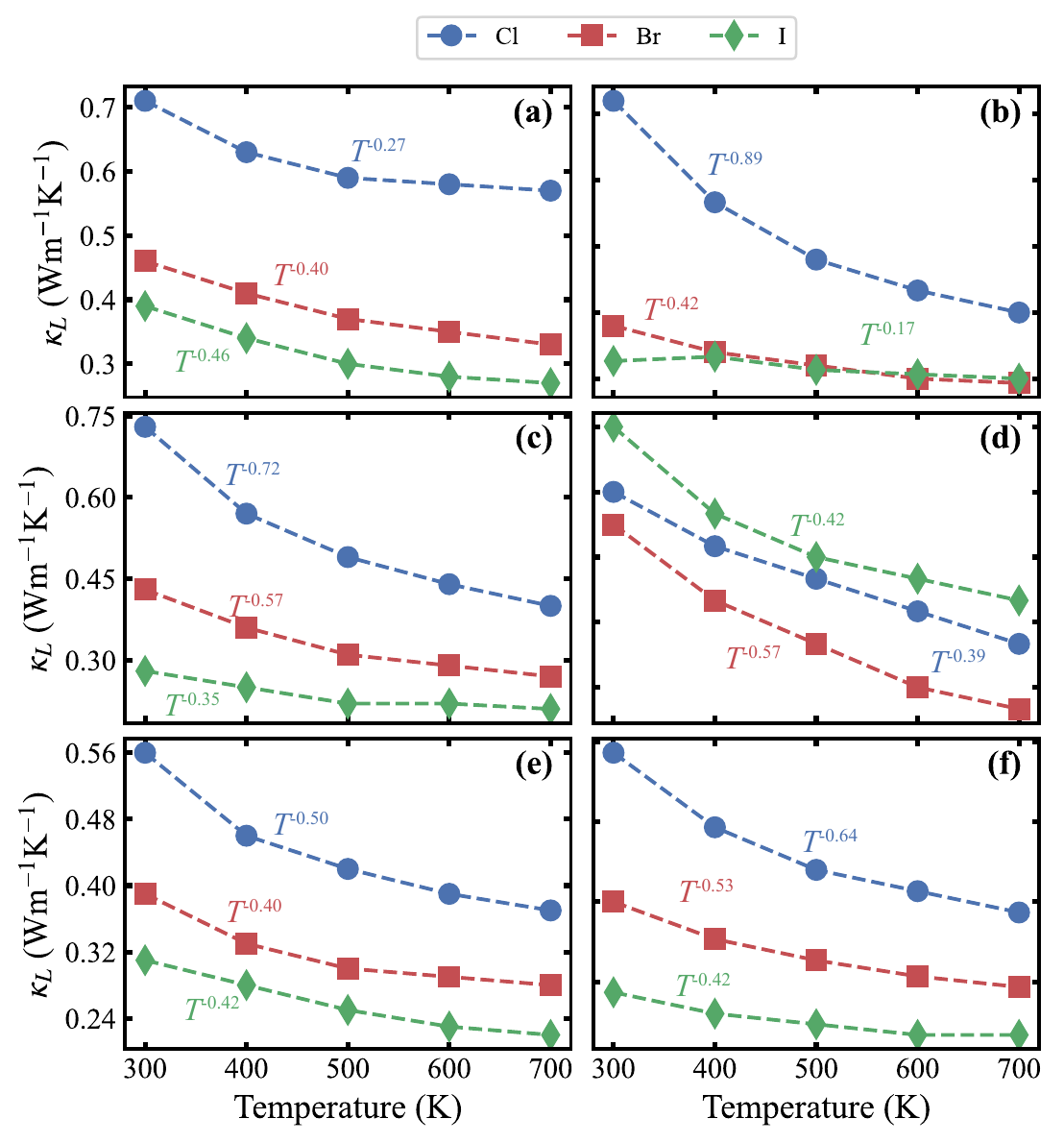}
	\caption{Temperature dependence of the lattice thermal conductivity $\kappa_{\mathrm{L}}$ of Cs$_2BX_6$ with $B=$ Zr, Pd, Sn, Te, Hf, and Pt in panels (a)–(f), respectively.}
	\label{fig:kappa-all}
\end{figure}

The temperature-dependent total lattice thermal conductivity ($\kappa_{\mathrm{L}}$), including contributions from $\kappa_{\mathrm{3,4ph}}^{\mathrm{SCPH}}$ and $\kappa_{\mathrm{3,4ph}}^{\mathrm{C}}$, is shown in Fig.~\ref{fig:kappa-all} over the temperature range of 300--700~K. All compounds exhibit intrinsically low $\kappa_{\mathrm{L}}$ values and a weak temperature dependence, which can be described by $\kappa_{\mathrm{L}} \propto T^{-\alpha}$ with $\alpha \ll 1$. This behavior arises from the combined effects of temperature-induced phonon frequency upshifts and an enhanced contribution from coherent phonon transport at elevated temperatures.

As shown in Fig.~\textcolor{magenta}{S7}, the frequencies of acoustic and low-lying optical phonons increase slightly with increasing temperature. This frequency hardening partially compensates for the enhanced phonon--phonon scattering at higher temperatures, leading to a substantial deviation from the $T^{-1}$ dependence expected for Umklapp-dominated thermal transport~\cite{peierls1996quantum,ziman2001electrons}. In addition, the coherent contribution to $\kappa_{\mathrm{L}}$ increases approximately linearly with temperature. Although its absolute magnitude remains relatively small for all compounds, this contribution counteracts the conventional decrease of $\kappa_{\mathrm{L}}$ with increasing temperature.

The temperature dependence of $\kappa_{\mathrm{L}}$ varies significantly with both the $B$- and $X$-site ions. In general, the exponent $\alpha$ for Cs$_2BX_6$ compounds decreases systematically from Cl to I, reflecting progressively weaker chemical bonding and, consequently, stronger lattice anharmonicity in the iodide compounds. For example, Cs$_2$PdI$_6$ exhibits the weakest temperature dependence ($\alpha = 0.17$), whereas Cs$_2$PdCl$_6$ shows the strongest dependence ($\alpha = 0.89$). An exception is found in the Cs$_2$Zr$X_6$ series, where the smallest and largest values of $\alpha$ occur in Cs$_2$ZrCl$_6$ and Cs$_2$ZrI$_6$, respectively, due to the stronger temperature dependence of $\kappa_{\mathrm{3,4ph}}^{\mathrm{C}}$ in Cs$_2$ZrCl$_6$.

\vspace{0.5cm}

\section{Conclusions}
In summary, we have performed a comprehensive first-principles study of the anharmonic lattice dynamics and lattice thermal conductivity of lead-free vacancy-ordered halide double perovskites Cs$_2BX_6$ ($B$ = Zr, Pd, Sn, Te, Hf, and Pt; $X$ = Cl, Br, and I). By combining the self-consistent phonon approach with a unified theory of thermal transport, we explicitly accounted for temperature-dependent phonon renormalization, three-phonon and four-phonon scattering, and coherent phonon contributions. Our calculations demonstrate that all compounds exhibit intrinsically ultralow lattice thermal conductivity, with values below 1.0~W\,m$^{-1}$\,K$^{-1}$ at 300~K. At room temperature, three-phonon scattering dominates the thermal resistance, while four-phonon scattering and coherent transport contribute less than 30\% and 20\%, respectively. Contrary to the commonly assumed rattling mechanism associated with large structural voids, our analysis combining with machine-learning-assisted approaches show that the ultralow $\kappa_{\mathrm{L}}$ in most Cs$_2BX_6$ compounds primarily originates from intrinsically weak chemical bonding in halide frameworks, leading to low sound velocities (1100--1600~m\,s$^{-1}$), rather than exceptionally strong anharmonic rattling of Cs atoms. An important exception is Cs$_2$SnI$_6$, where the large phonon scattering phase space associated with the localized vibration of I atom and strong three-phonon scattering in the low-frequency range result in the lowest lattice thermal conductivity among all studied compounds. Overall, this work provides a unified microscopic understanding of thermal transport in Cs$_2BX_6$ compounds and establishes fundamental design principles for developing electronic insulators and functional materials with ultralow lattice thermal conductivity.

\section{ACKNOWLEDGMENTS}
The authors acknowledge the support received from the National Natural Science Foundation of China (Grant No. 12374024) and State Key Laboratory for Advanced Metals and Materials, Grant No. 2025-Z15.

\bibliography{ref}
\end{document}